\documentclass[apjl]{emulateapj}

\usepackage{url}
\usepackage{color}

\shorttitle{ALMA Science Verification Data: 3C~286}
\shortauthors{Nagai et al.}

\begin{document}


\title{ALMA Science Verification Data: Millimeter Continuum Polarimetry of the Bright Radio Quasar 3C~286 }


\author{H. Nagai\altaffilmark{1}, K. Nakanishi\altaffilmark{1,2,6}, R. Paladino\altaffilmark{3}, C. L. H. Hull\altaffilmark{4,5}, P. Cortes\altaffilmark{6,8}, G. Moellenbrock\altaffilmark{7}, E. Fomalont\altaffilmark{6,8}, K. Asada\altaffilmark{9}, \& K. Hada\altaffilmark{1}}
%\affil{National Astronomical Observatory of Japan}
\email{hiroshi.nagai@nao.ac.jp}

%\and


\altaffiltext{1}{National Astronomical Observatory of Japan, Osawa 2-21-1, Mitaka, Tokyo 181-8588, Japan}
\altaffiltext{2}{The Graduate University for Advanced Studies (SOUKENDAI), Osawa 2-21-1, Mitaka, Tokyo 181-8588, Japan}
\altaffiltext{3}{INAF-Osservatorio di Radioastronomia, Via P. Gobetti, 101 40129 Bologna, Italy}
\altaffiltext{4}{Harvard-Smithsonian Center for Astrophysics, 60 Garden St., Cambridge, MA 02138, USA}
\altaffiltext{5}{Jansky Fellow of the National Radio Astronomy Observatory}
\altaffiltext{6}{Joint ALMA Observatory, Alonso de C\'ordova 3107, Vitacura 763 0355, Santiago de Chile}
\altaffiltext{7}{National Radio Astronomy Observatory, Socorro, NM 87801, USA}
\altaffiltext{8}{National Radio Astronomy Observatory, Charlottesville, VA 22903-2475, USA}
\altaffiltext{9}{The Academia Sinica Institute of Astronomy and Astrophysics, AS/NTU. No.1, Sec. 4, Roosevelt Rd, Taipei 10617, Taiwan, R.O.C.}

\begin{abstract}
\noindent
We present full-polarization observations of the compact, steep-spectrum radio quasar 3C~286 made with the ALMA at 1.3~mm.  These are the first full-polarization ALMA observations, which were obtained in the framework of Science Verification.  A bright core and a south-west component are detected in the total intensity image, similar to previous centimeter images.  Polarized emission is also detected toward both components.  The fractional polarization of the core is about 17\%; this is higher than the fractional polarization at centimeter wavelengths, suggesting that the magnetic field is even more ordered in the millimeter radio core than it is further downstream in the jet.  The observed polarization position angle (or EVPA) in the core is $\sim$\,$39^{\circ}$, which confirms the trend that the EVPA slowly increases from centimeter to millimeter wavelengths.  With the aid of multi-frequency VLBI observations, we argue that this EVPA change is associated with the frequency-dependent core position. We also report a serendipitous detection of a sub-mJy source in the field of view, which is likely to be a submillimeter galaxy.
\end{abstract}		



\section{Introduction}
3C~286 is a bright radio quasar at the redshift $z = 0.849$ \citep{Burbidge1969}.  The radio source shows a steep spectrum at gigahertz frequencies with the spectral peak at about 100~MHz.  The arcsecond-scale radio structure is dominated by a compact core with some extended structures \citep{Akujor1995}.  A jet-like feature extends $\sim3\arcsec$ from the bright central component at a position angle of $\sim-115^{\circ}$ with an extended lobe. There is another jet-like feature at a position angle of $\sim90^{\circ}$, which could be a counter jet component.  The projected distance of the radio emission is about 15~kpc, and 3C~286 is classified as a compact steep spectrum (CSS) radio sources \citep[e.g.,][]{Fanti1985, O'Dea1998}.  

3C~286 is remarkably highly polarized and its polarization properties are extremely stable \citep{Perley2013}.  The fractional polarization is about 10\% at 20~cm and increases with increasing frequency \citep{Perley1982}.  The electric vector position angle (EVPA) of the polarization is about $33^{\circ}$ at 20~cm  \citep{Perley1982}, which is roughly aligned with the direction of south-west jet-like feature.  It is known that the EVPA has no or little frequency dependence at centimeter wavelengths  \citep{Perley1982, vanBreugel1984}.  This source is one of the main polarization calibrators used for Very Large Array (VLA) observations at a wide range of wavelengths and Very Long Baseline Array (VLBA) at wavelengths longer than $\sim6$~cm.  More recently, the polarization properties were also measured at millimeter wavelengths using the VLA, the IRAM 30\,m telescope, and CARMA \citep{Agudo2012, Perley2013, Hull2015}; it is evident that the EVPA increases slowly up to $\sim38^{\circ}$ at 1.3~mm.  

The Atacama Large Millimeter and Submillimeter Array (ALMA) released the Science Verification (SV) data of 3C~286 on 2015 July 28.  These are the first full-polarization observations performed by ALMA, and are useful for demonstrating the capability of the ALMA polarization system.  In this paper we present the results from the SV data, as well as an interpretation of the frequency dependence of the EVPA and the nature of 3C~286 with the aid of VLBA data.  We also give a detailed characterization of the performance of the ALMA polarization system in Appendix A.1.

\section{Observation and Data Reduction}
\label{sec:obs}
The observations were performed on 2014 July 1 and July 2 using thirty-one 12\,m antennas in Band 6 (1.3~mm). Both orthogonal linear polarizations ($X$ and $Y$) were received simultaneously and the data were correlated using the ALMA 64-input correlator to obtain $XX$, $YY$, $XY$, and $YX$ correlations.  Two spectral windows (spws) were set in both the lower sideband (LSB) and the upper sideband (USB), with 64 channels per spw and 31.25~MHz channel resolution, providing a bandwidth of 2~GHz per spw.  The central frequencies in spws 0, 1, 2, and 3 are 224~GHz, 226~GHz, 240~GHz, and 242~GHz, respectively.  The observations of 3C~286 comprise multiple scans with 7\,min integrations, yielding $\sim$\,2 hours of integration in total.  3C~279, Ceres, and J1310+3220 were observed as the bandpass, flux, and gain calibrators, respectively.  In addition, J1337-1257 was observed every $\sim$\,30 minutes for calibration of the instrumental polarization ($D$-terms), cross-hand delay, and cross-hand phase.  

The data reduction was done using Common Astronomical Software Applications (CASA) version 4.3.1.  Amplitude calibration was performed using measurements of the system temperature ($T_\textrm{sys}$) on a channel-by-channel basis.  Rapid phase variations on timescales of less than the gain calibration cycle were corrected using the water vapor radiometer.  In addition to unreliable data such as amplitude/phase jumps and low antenna gains, several channels at both edges of each spw were flagged because of low power caused by filtering effects.  The bandpass calibration was done both in phase and amplitude. Temporal variation of the gain amplitude and phase was calibrated using the averaged $XX$ and $YY$ correlations of a gain calibrator.  
  
To calibrate the instrumental polarization, we first obtained the gain calibration solution without any source polarization model.  Such a gain solution absorbs the source polarization.  To extract the Stokes $Q$ and $U$ of the calibrator hidden in the gain solution, we used the function \texttt{qufromgain} in an add-on script \texttt{almapolhelpers.py} in CASA.  The cross-hand delay and cross-hand phase differences relative to a reference antenna were calibrated using the CASA task \texttt{gaincal} with the modes \texttt{KCROSS} and \texttt{XYf+QU} respectively.  After the cross-hand delay and phase calibration, we performed the instrumental polarization calibration using CASA task \texttt{polcal}.  The derived instrumental polarization level is typically 2--4\%, but several antennas showed more than 5\% at some channels.  Significant frequency structure is found on all antennas.  For a consistency check of instrumental the polarization calibration we performed the polarization calibration using 3C~279, which was originally intended for bandpass calibration.  The resulting instrumental polarization obtained from 3C~279 showed a good agreement with that from J1337-1257.  The instrumental polarization properties are discussed in more detail in Appendix A.1.  The flux scaling factor was estimated using Ceres.  The flux densities of 3C~279 and J1310+3220 were reasonably consistent with the flux measurements in ALMA database.  The data reduction procedure presented here is also described in the CASA guide.\footnote{3C~286 polarization CASA guide: \url{https://casaguides.nrao.edu/index.php/3C286\_Polarization}}  

In addition to performing the calibration in the CASA guide, we refined the gain calibration. After applying all calibration was applied we found that gain calibrator was strongly polarized.  However, in the gain calibration process we assumed the source was unpolarized; this could cause an error in gain calibration.  To avoid this, we constructed Stokes $I$, $Q$, and $U$ models of the gain calibrator using the \texttt{CLEAN} algorithm.  After the model construction, we re-performed the gain calibration of the gain calibrator based on the models, and then applied the calibration to 3C~286.

After all calibrations were applied, we still found significant calibration residuals in both the phase and amplitude of 3C~286 as a function of time, particularly in the first half of observation.  We excluded those data before making images.  To correct the calibration residuals in the rest of the data we used the self-calibration technique.  First, we constructed Stokes $I$, $Q$, and $U$ models of 3C~286 using \texttt{CLEAN} with Briggs weighting and a robustness value of 0.5.  Next, using the \texttt{CLEAN} model, we performed phase self-calibration with a solution interval of 30 seconds.  Finally, we performed both amplitude and phase self-calibration (\texttt{calmode=ap}) with a solution interval of 30 seconds.  After all of these processes the time dependence of phase and amplitude were significantly improved.  We obtained a final image rms of 0.05~mJy beam$^{-1}$, 0.04~mJy beam$^{-1}$, and 0.04~mJy beam$^{-1}$ for Stokes $I$, $Q$, and $U$, respectively.  Those are slightly larger than the values reported in the 3C~286 CASA guide because we used fewer data for the imaging.  The theoretical thermal noise can be estimated to be about 0.02~mJy beam$^{-1}$ with twenty-nine antennas, $T_\textrm{sys}\simeq100$~K, and an integration time of 1 hour.  Thus, the Stokes $I$ image is dynamic range limited.  The polarization sensitivity is probably limited by the error in the instrumental polarization calibration.  This is discussed in Appendix A.1. %Note that Stokes V emission was apparently seen on the image, but this is resulted from the artifact that we assumed Stokes V of J1337-1257 is zero.

Since the absolute feed position orientation is known from the front-end assembly process\footnote{The polarization X of receivers are aligned radially in the receiver cryostat and toward the cryostat center. The polarization Y is perpendicular to X.  ALMA Band 6 feeds are tilted at $45^{\circ}$ to the X and Y axes of the antenna coordinate. See ALMA Technical Handbook for more detail (\url{https://almascience.eso.org/proposing/call-for-proposals/technical-handbook}).}, we do not need to observe a source with a known polarization angle for the absolute EVPA calibration in contrast to the telescope with circular polarized feeds (e.g., VLA). The major cause of EVPA error is probably the misalignment of front-end feeds.  The misalignment of the absolute position angle would not be larger than 2$^{\circ}$ (priv. comm. with S. Asayama), which is also defined in the ALMA Subsystem Technical Requirements.  If we assume that the amount of misalignment is all the same but the direction is random over all antennas, the EVPA measurement accuracy ($\Delta\chi$) in the synthesized image would be $\Delta\chi\sim \psi/\sqrt{N}$ where $\psi$ is the amount of misalignment and $N$ is the number of antennas.  For this dataset, two antennas were completely flagged out because of anomalous amplitude, thus $\Delta\chi<2^{\circ}/\sqrt{29}=0.4^{\circ}$.  Throughout this paper, we adopt $0.4^{\circ}$ for the absolute EVPA accuracy.  
		
\section{Results}
\begin{figure}
\begin{center}
\includegraphics[width=8.5cm]{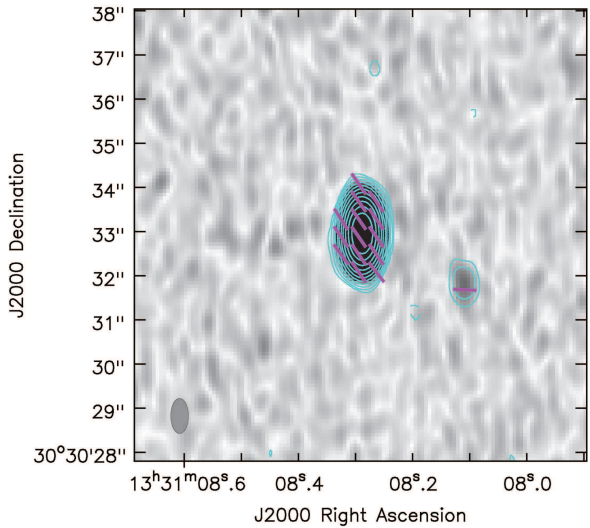}
\end{center}
\caption{Cyan contours show Stokes $I$ intensity, grayscale shows polarized intensity ($\sqrt{Q^{2}+U^{2}}$), and magenta vectors show the polarization position angle of 3C~286.  The contours are plotted at (--1, 1, 2, 4, 8, 16, 32, 64, 128, 512) $\times$0.15~mJy beam$^{-1}$, which is $3 \times$ the rms noise in the Stokes $I$ image.  The EVPA is plotted at levels $>3\sigma \times$ the rms noise in the polarized intensity map.  The restoring beam is $0.797\arcsec \times 0.406\arcsec$ at a position angle of $-180^{\circ}$, which is given by the ellipse at the bottom left corner of the image.}
\label{3C286map}
\end{figure}

Figure \ref{3C286map} shows a polarized intensity map of 3C~286, with EVPA overlaid on Stokes $I$ contours.  In the Stokes $I$ image both the bright core and the south-west (SW) jet are clearly detected.  The flux densities of the core and the SW component are summarized in Table \ref{table_flux}.  The SW component is detected for the first time at this frequency.  The flux density of the SW component reported in \cite{Akujor1995} is 31.4~mJy at 8.4~GHz.  The spectral index between 8.4~GHz and 230~GHz is estimated to be about $-1$.  Non detection of the counter jet component gives an upper limit of spectral index $\alpha\lesssim -1.4$ between 8.4~GHz and 230~GHz. 

\begin{figure}
\begin{center}
\includegraphics[width=8.5cm]{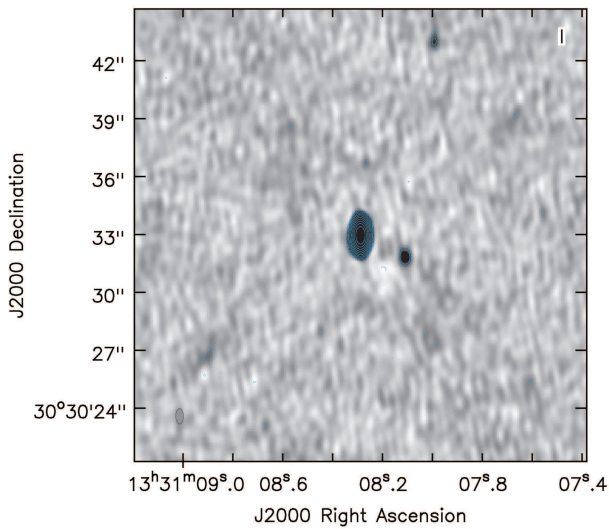}
\end{center}
\caption{$24\arcsec\times 24\arcsec$ image of Stokes $I$ intensity centered at the core of 3C~286.  Both contour and grayscale indicate Stokes $I$ intensity.  The contours are plotted at the same level as in Figure \ref{3C286map}.}
\label{3C286WF}
\end{figure}

\begin{table}
\caption{Peak intensity of the core and the SW component}\label{table_flux}
  \begin{center}
    \begin{tabular}{cccc} \hline\hline
Component & I\tablenotemark{a} & P\tablenotemark{b} & F\tablenotemark{c} \\ \hline
Core & 387.77$\pm$0.05~mJy & 64.81$\pm$0.07~mJy & 0.17$\pm$0.02\ \\
SW & 0.55$\pm$0.05~mJy & 0.21$\pm$0.07~mJy & 0.38$\pm$0.12\ \\ \hline
   \end{tabular}	
  \end{center}
\tablenotetext{1}{Peak intensity [mJy beam$^{-1}$]}
\tablenotetext{2}{Polarized peak intensity [mJy beam$^{-1}$]}
\tablenotetext{3}{Fractional polarization}
 \end{table}		

Polarized emission is clearly detected toward the core and is also marginally detected toward the SW component.  The polarization percentage of the core is 16.7$\pm$0.2\%, which is higher than at longer wavelengths.  Since the signal-to-noise ratio of the SW component is low, we should take into account the polarization bias \citep{Vaillancourt2006} for this component.  The bias-corrected polarization intensity of SW component is estimated to be 0.2~mJy beam$^{-1}$, which corresponds to a $2.8\sigma$ detection.  The EVPA of the core is $38.6\pm0.4$~degrees.  Note that the statistical error of the EVPA is $\sim$\,$10^{-2}$ degrees, which is negligible compared with the systematic error.  The obtained EVPA shows a good agreement with \cite{Hull2015} and \cite{Agudo2012}, but our ALMA measurement of fractional polarization and EVPA has better accuracy thanks to ALMA's excellent sensitivity and instrumental polarization calibration accuracy.  The polarization percentage of the SW component is larger than 30\%, which indicates an even more ordered magnetic field than in the core.  The EVPA of the SW component is $\sim$\,$90^{\circ}$, which is consistent with results from longer wavelength observations \citep{Akujor1995}.

We also report the serendipitous detection of a point source.  Figure \ref{3C286WF} shows a wide field image ($24\arcsec \times 24\arcsec$) centered on the core of 3C~286.  A $6\sigma$ point source is detected at the position (RA, DEC)=(13:31:08.0, 30:30:43.0), $\sim11\arcsec$ NNW
%($\sim10.0\arcsec$Ein RA and $\sim3.8\arcsec$Ein DEC) 
of the core of 3C~286.
%at the position angle of about $-20^{\circ}$.  
The peak intensity is about 0.5~mJy beam$^{-1}$ after the primary beam correction.  Although the position was derived from the self-calibration image, we expect the position uncertainty to be $<0.1\arcsec$ since the measured position of 3C~286 core is consistent with the position derived by the VLBI astrometry \citep{Beasley2002} within 0.1$\arcsec$. No counterpart is found at this position in Hubble Space Telescope (HST) \citep{Brun1997} or Sloan Digital Sky Survey (SDSS) images.  This source is likely to be a submillimeter galaxy (SMG), one of a population of distant galaxies where star formation is obscured by the dust \citep[e.g.,][]{Hughes1998}.  The detection of a SMG is consistent with the source number counts of sub-mJy sources in the SXDF-ALMA deep survey \citep{Hatsukade2016}.  Although centimeter data is also important to conclude whether this is a true SMG, it is difficult to locate a centimeter counterpart in the image from the FIRST survey \citep{Becker1995} because of the contamination by the bright 3C~286 emission.  

\section{Discussion} 
The ALMA SV data presented here confirm that the EVPA increases slowly from centimeter to millimeter wavelengths.   This trend cannot be explained by Faraday rotation since the change in the EVPA is not observed in centimeter bands \citep[$\lambda\gtrsim10$~cm,][]{Perley2013}.  It is thus reasonable to assume that this change in EVPA is related to the frequency-dependent location of the brightest spot along the jet.  In general, synchrotron opacity becomes higher when approaching the base of a quasar jet; thus, the inner part of the jet becomes optically thinner at shorter wavelengths.  The emission from extended jets also becomes fainter at shorter wavelengths, so the bulk of the emission we detect should be radiated from the inner part of the jet.  The observed frequency dependence of the EVPA probably suggests that the magnetic field in the inner part of the jet is slightly misaligned with the field in the outer part.  This possibility was also pointed out by \cite{Agudo2012}.  

\begin{figure}
\begin{center}
\includegraphics[width=7cm]{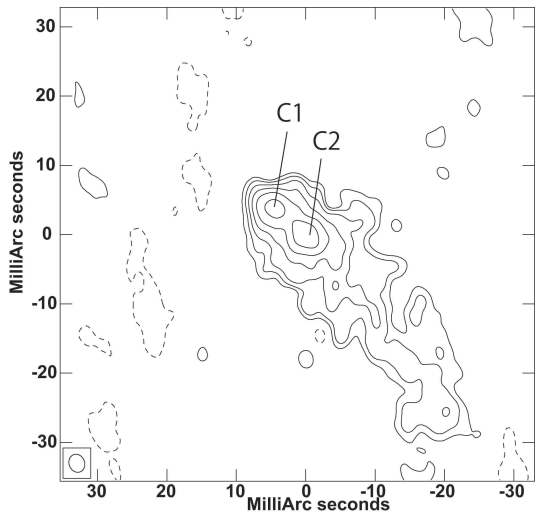}
\end{center}
\caption{Contour plot of the VLBA image of 3C~286 at 4~cm on 1997 January 11.  The contours are plotted at ($-1$, 1, 2, 4, $\dots$, 32) $\times$ 4.9~mJy beam$^{-1}$. The peak intensity is 284.7~mJy beam$^{-1}$. The convolved beam size is ($2.65 \times 2.24$)~mas at a position angle of $21.2^{\circ}$. }  
\label{fig:3C286VLBA}
\end{figure}

For a more quantitative discussion, we analyzed archival VLBA images at 13~cm, 4~cm, and 2~cm obtained from the ``Monitoring Of Jets in Active galactic nuclei with the VLBA Experiments'' (MOJAVE) and the Radio Reference Frame Image Database (RRFID).  The summary of data is listed in Table \ref{tab:separation}.  Figure \ref{fig:3C286VLBA} shows the total intensity image at 4~cm.  Two bright knots (C1 and C2) in the central region and extended cocoon-like emission are detected.  We performed Gaussian model fitting to C1 and C2 using the AIPS task \texttt{JMFIT}; the derived physical parameters are summarized in Table \ref{tab:separation}.  The integrated fluxes of C1 and C2 derived by this fitting are not very consistent from epoch to epoch, probably because of different \textit{uv}-coverages.  However, the derived position is more or less consistent.  Thus, we show only the position information in this table.  Note that the absolute position of each component is lost by fringe fitting and self-calibration.

\begin{table*}
\caption{Separation between C1 and C2}\label{tab:separation}
  \begin{center}
    \begin{tabular}{cccccc} \hline\hline
Wavelength & Frequency & Epoch & Ref.\tablenotemark{a}  & Separation in RA [mas] \tablenotemark{b} & Separation in DEC [mas] \tablenotemark{c}  \\ \hline
13~cm & 2.32~GHz & 1997 January 11 & (2) & 4.02 & 3.33 \\  
13~cm & 2.29~GHz & 2000 October 23 & (1) & 3.87 & 2.86 \\ 
4~cm & 8.55~GHz & 1997 January 11 & (2) & 4.34 & 3.72 \\
2~cm & 15.35~GHz & 1995 April 07 & (3) & 4.92 & 4.06 \\
2~cm & 15.34~GHz & 1997 March 10 & (3) & 4.76 & 3.94 \\
2~cm & 15.34~GHz & 2002 August 11 & (3) & 4.91 & 4.06 \\
\hline
   \end{tabular}	
  \end{center}
\tablenotetext{1}{References: (1) Unpublished RRFID data, (2) RRFID data (Fey et al. 2000), (3) MOJAVE data (Lister et al. 2005)}
\tablenotetext{2}{Separation of C1 and C2 in right ascension}
\tablenotetext{3}{Separation of C1 and C2 in declination}
 \end{table*}	

The possibility of a frequency-dependent position of the bulk emission can be tested if the frequency-dependent position of the radio core---the so-called core shift effect---is observed \citep[e.g.,][]{Hada2011}.  We thus measured the relative distance between C1 and C2 at various frequencies.  
Assuming that C2 has no frequency dependence of its position, the change in the relative distance can be regarded as the position shift of the radio core (C1).  

Figure \ref{fig:RelativePosition} shows the position of C1 relative to C2 (see the Appendix for details).  The positional uncertainty due to the fitting error is smaller than the size of each symbol, typically $<0.1$\%.  Although the measured position at the same frequency has an offset with a level of a few tenth milli-arcsec (mas) among measurements in different epochs, there is a general trend that the position is shifted to the upstream of the jet with increasing frequency.  The shift occurs at a position angle of about $45^{\circ}$, which is consistent with the jet position angle.  Even though the data were taken non-simultaneously, the position shift of the core due to the emergence of a new component and/or a change in the opacity should not be relevant since the radio flux density of 3C~286 has remained unchanged for decades.  There is no monotonic increase in the separation with time among three datasets at 2~cm, which indicates that there is no significant motion of C2.  Thus, the difference in measured position between epochs can be regarded as the uncertainty due to the systematic error, most likely from different \textit{ uv}-coverages between datasets.

\begin{figure}
\begin{center}
\includegraphics[width=7cm]{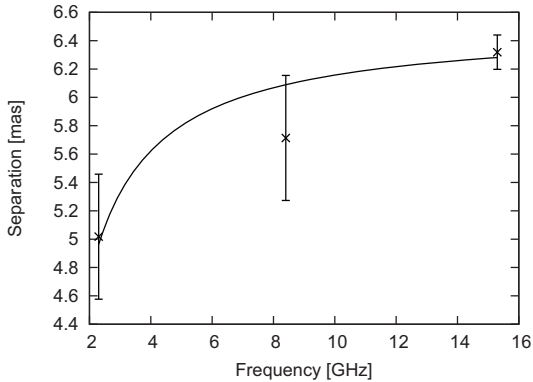}
\end{center}
\caption{The separations between C1 (core) and C2 as a function of frequency.  The curved line is the best-fit power law: $r_\textrm{c}~=~(6.51\pm0.17) +(-3.57\pm1.13)\,\nu^{-1}$.}
\label{fig:coreshift}
\end{figure}

Figure \ref{fig:coreshift} shows the separation between C1 and C2 as a function of frequency.  At 2~cm and 13~cm, we use the average value of the various epochs.  The errors at 2~cm and 13~cm are set to be the maximum difference in measured positions (projected to the jet position angle) between different epochs.  Since we have only one measurement at 4~cm, this method is not applicable, but it is naturally expected that the error is approximately proportional to the observed wavelength.  We conservatively adopt the same error for the 13~cm and 4~cm data. Given that the core (C1) has a flat-spectrum in centimeter bands \citep{Zhang1994, Cotton1997}, it is expected that the core position $r_{c}(\nu)$ is a power-law function of frequency \citep[$r_\textrm{c}\propto\nu^{-1/k}$, e.g.,][]{Bartel1986, Lobanov1998}.  Since we have only three data points, it is not possible to perform a fit with the function $r_\textrm{c}=A+B\nu^{-1/k}$, leaving $A$, $B$, and $k$ as free parameters.  Instead, we assume $k=1$, which is the case for a conical jet \citep{Konigl1981} and generally provides a good fit for most quasar jets \citep{Sokolovsky2011}.  Using the power-law function to extrapolate the relative distance at 230~GHz (1.3~mm), we expect the 1.3~mm radio core to be located $\sim$\,0.17 mas upstream of the 2~cm core.  This corresponds to $\sim$\,1.3~pc in projected distance.  Thus, the observed change in the polarization angle from centimeter to millimeter wavelengths can be attributed a change in the magnetic field direction over a distance of $\sim$\,1\,pc.  

Above we assumed that C2 is optically thin and that the position of C2 has little to no frequency dependence.  This is valid for most non-core components in radio jets at centimeter wavelengths \citep{Sokolovsky2011}.  However, it is reported that C2 also has a flat or inverted (optically thick) spectrum between 5~GHz and 15~GHz \citep{Zhang1994, Cotton1997}, although flux measurements are somewhat uncertain due to complex source structure imaged with different \textit{ uv}-coverages.  We also tried to measure the flux density by using \texttt{JMFIT} to derive the spectral index, but the result is not very consistent between different epochs for the same reason.  If C2 is optically thick, it is possible that the measured frequency dependence of the separation is not due to a core-shift effect.  However, there are two pieces of evidence that lead us to suspect that C1 is more optically thick than C2, and thus that the observed separation change is dominated by the core shift.  One is that the fractional polarization of C1 is slightly smaller than that that of C2 \citep{Jiang1996}, which is consistent with the properties of polarization in the optically thick regime \citep{Jones1977}.  The other reason is that C1 is more compact than C2 at 1.7~GHz and 5~GHz, as reported in previous papers \citep{Zhang1994, Cotton1997}.  Since C1 and C2 show more or less the same flux density \citep{Zhang1994, Cotton1997}, C1 should have a higher brightness temperature, and thus a higher optical depth. 

\begin{figure}
\begin{center}
\includegraphics[width=7cm]{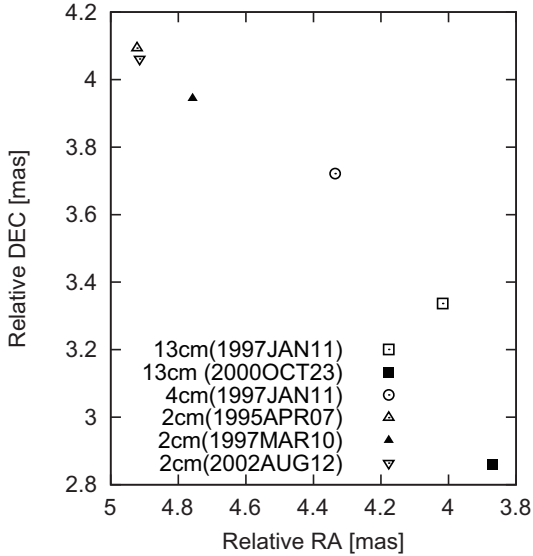}
\end{center}
\caption{Relative positions of C1 (core) and C2 at different frequencies and different epochs.}  
\label{fig:RelativePosition}
\end{figure}

Alternatively, the separation change between C1 and C2 can happen if C1 and C2 are comprised of multiple components and these components have different spectra.  Also, the change in fractional polarization and EVPA can result from a slightly different spectral index and intrinsic polarization between C1 and C2.  Therefore, the core-shift could not be the only mechanism to tie the change in the polarization properties.  Localizing the dominant polarized component with millimeter VLBI observations would give a conclusive answer.

\cite{Cotton1997} proposed a bent-jet model in which the jet direction of 3C~286 is initially misaligned and then curves toward the observer.  \cite{Agudo2012} also pointed out that the frequency dependence of the observed EVPA could be accommodated with the bent-jet model.  A bent jet is could indeed explain the frequency-dependent EVPA; however, in our view, it is unlikely that the jet emission from 3C~286 is beamed to the observer as proposed by \cite{Cotton1997}, since 3C~286 shows a very little flux variation.  More moderate jet bending seems to fit into the observed 3C~286 properties.  

The trend of higher fractional polarization at shorter wavelength suggests that the magnetic field is more ordered toward the inner region of the jet.  Since significant Faraday rotation is not observed on 3C~286, the observed polarization angle indicates that the projected magnetic field is perpendicular to the jet axis. Perpendicular fields can result from shock compression \citep{Laing1981}.  However, the perpendicular fields continue up to 50~mas (375~pc in projected distance) from the core \citep{Cotton1997, Jiang1996} in 3C~286, and it is unlikely that there would be a continuous and stable shock over this distance.  An alternative is that 3C~286 might have toroidally-dominated helical magnetic field, which is predicted by the Blandford-Znajek mechanism \citep[e.g.,][]{Blandford1977} and recent magneto-hydrodynamical (MHD) models \citep[e.g.,][]{McKinney2006}.

\section{Conclusion}
We have used ALMA polarization SV to produce polarization image of the CSS radio quasar 3C~286 at 1.3~mm with a very high accuracy.  The core is polarized at the level of $\sim$\,17\% and the EVPA is $\sim$\,39$^{\circ}$; previous observations at centimeter wavelengths reported fractional polarization and EVPA values of $\sim$\,10\% and $\sim$\,$33^{\circ}$, respectively.  If the observed frequency dependence of the separation between C1 and C2 revealed by VLBI observations indicates the core-shift effect, then the millimeter emission must come from the inner part of the jet.  Thus, overall results suggest both that the millimeter radio core has an even more ordered magnetic field than the centimeter radio core, and that the magnetic field orientation changes by about $6^{\circ}$ over the distance between centimeter and millimeter cores.
	
\bigskip
HN thanks \'Alvaro Gonz\'alez and Shinichiro Asayama for the useful discussion about laboratory measurements of the instrumental polarization.  This paper makes use of the following ALMA data: ADS/JAO.ALMA\#2011.0.00017.SV. ALMA is a partnership of ESO (representing its member states), NSF (USA) and NINS (Japan), together with NRC (Canada), NSC and ASIAA (Taiwan), and KASI (Republic of Korea), in cooperation with the Republic of Chile. The Joint ALMA Observatory is operated by ESO, AUI/NRAO and NAOJ.  The National Radio Astronomy Observatory is a facility of the National Science Foundation operated under cooperative agreement by Associated Universities, Inc.  This research has made use of the NASA/IPAC Extragalactic Database (NED), which is operated by the Jet Propulsion Laboratory, California Institute of Technology, under contract with the National Aeronautics and Space Administration.  This research has made use of data from the MOJAVE database that is maintained by the MOJAVE team \citep{Lister2009}.  This research has made use of the United States Naval Observatory (USNO) Radio Reference Frame Image Database (RRFID).  HN is supported by MEXT KAKENHI Grant Number 15K17619.

\renewcommand{\bibname}{}

\appendix
\section{A.1 Polarization Calibration}  
\subsection{A.1.1 Instrumental Polarization}
\begin{figure}
\begin{center}
\includegraphics[width=16cm]{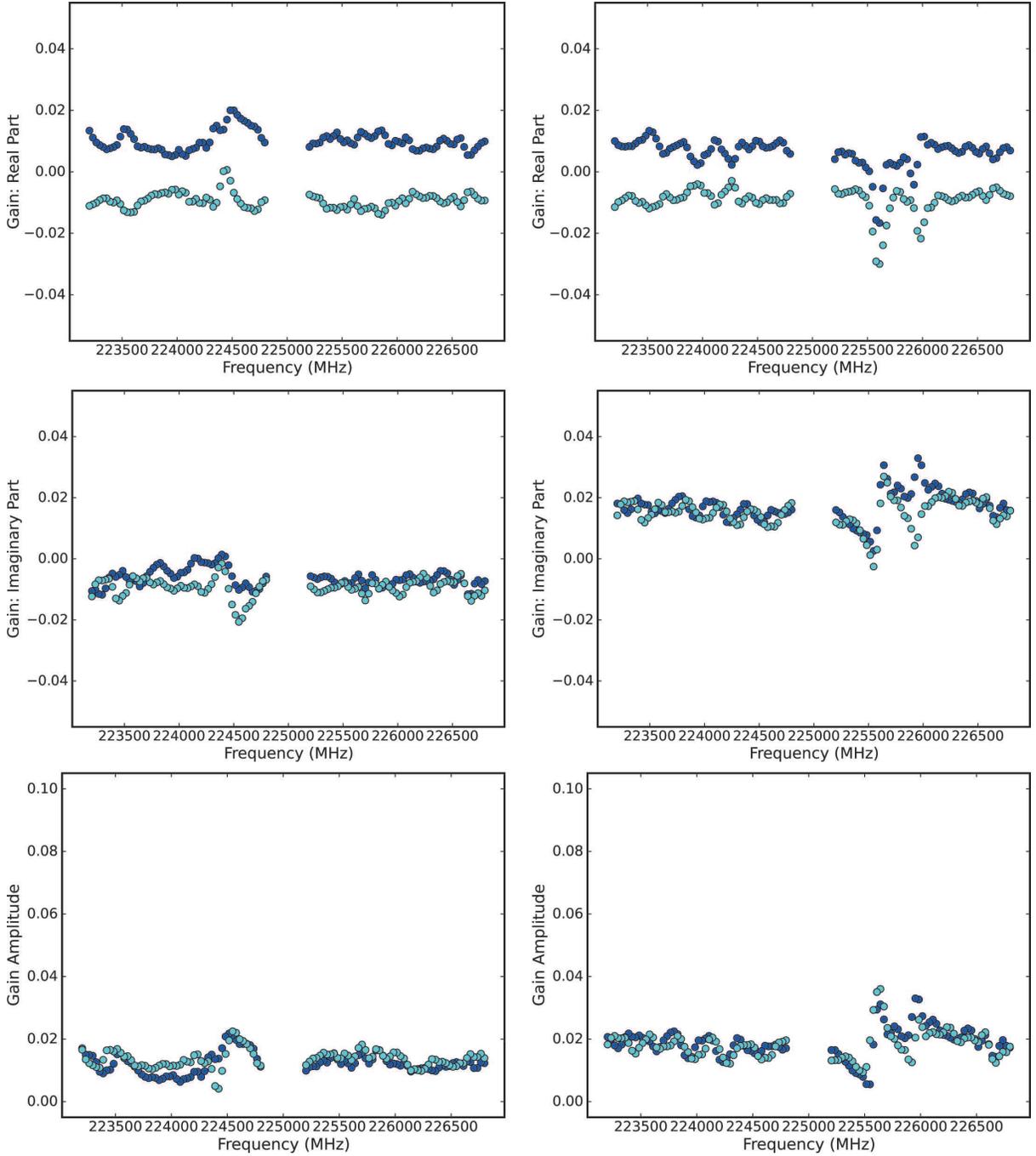}
\end{center}
\caption{Exaples of $D$-terms.  The left and right columns shows the $D$-term spectra for antennas DA41 and DV01, respectively.  The upper, middle, and lower panels show the real part, imaginary part, and amplitude of the $D$-terms. Blue and cyan symbols represent $D_\textrm{X}$ and $D_\textrm{Y}$.  Note that the $D$-terms presented in this figure were not computed in the telescope frame but in the sky frame; this solution can be obtained using the CASA command \texttt{dxy} in \texttt{almapolhelpers.py}. }  
\label{fig:Dplot}
\end{figure}

\begin{figure}
\begin{center}
\includegraphics[width=16cm]{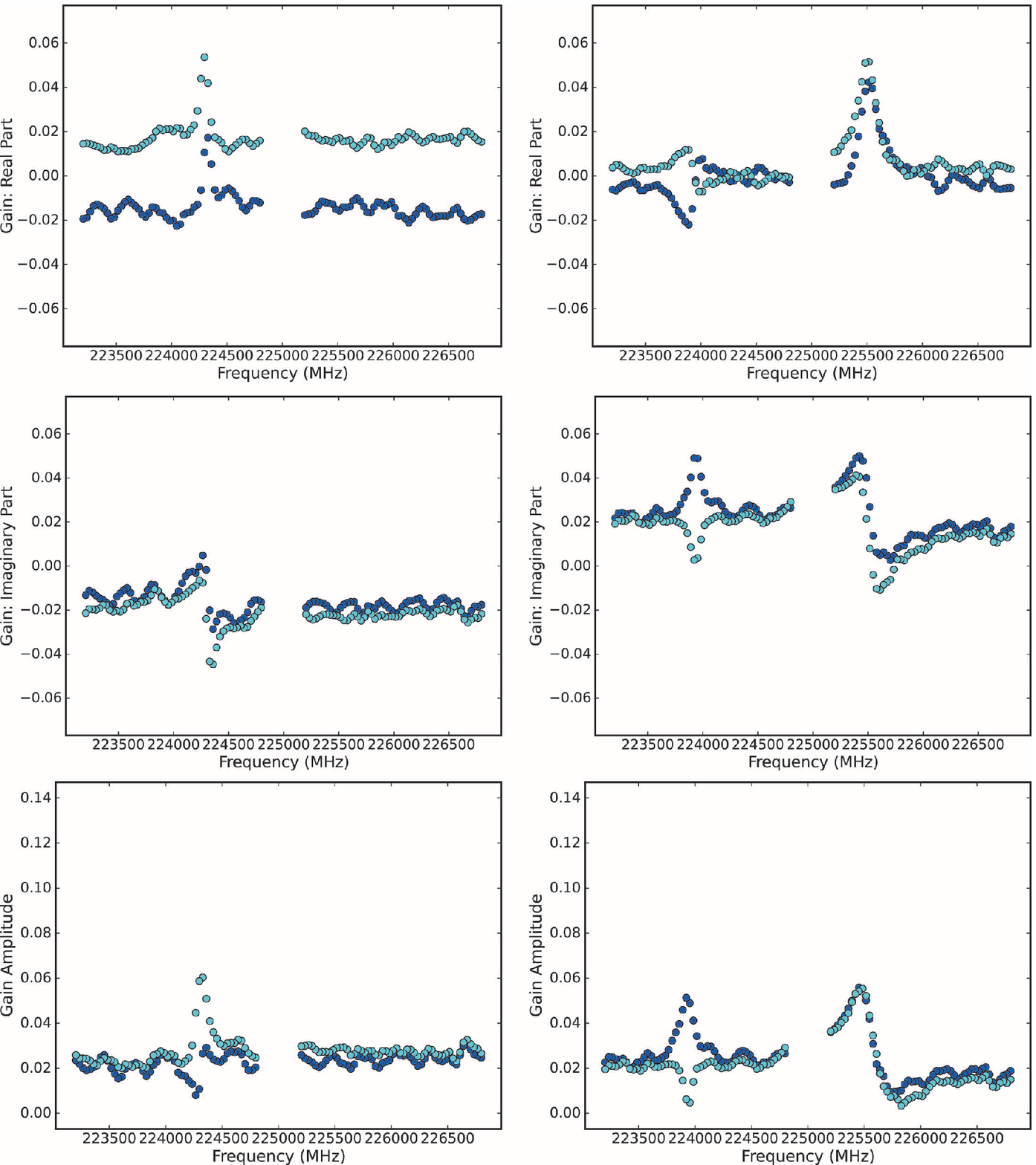}
\end{center}
\caption{The same plots as Figure \ref{fig:Dplot}, but for antennas showing large instrumental polarization (left: antenna DA54, right: antenna DV07).}
\label{fig:Dplot_bad}
\end{figure}

Figure \ref{fig:Dplot} shows some examples of instrumental polarization as a function of frequency.  Although antenna structure is different between the DV (North American) antennas and DA (European) antennas, the global properties of the instrumental polarization are similar.  Unfortunately, since no PM (East Asian) antennas participated to this observation, we were not able to check the instrumental polarization properties for PM antennas using this dataset.  However, we tested the instrumental polarization properties of PM antennas during commissioning and found no significant difference.  The instrumental polarization level is typically 2--4\%, but several antennas exhibit instrumental polarization levels >5\% at certain frequencies (see Figure \ref{fig:Dplot_bad}).  Significant frequency structure is seen in all antennas.  The frequency structure is quasi-periodic, and there is a resonance-like pattern at certain channels.  The frequency structure does not have exactly the same pattern among different antennas, but the periodicity looks similar among antennas.  Although there are no Band 6 laboratory measurements of the frequency dependence of instrumental polarization with such a high spectral resolution, the simulation and laboratory measurements of the Band 4 receivers hint at the origin of the frequency structure.  It is reported that the feed horn, infrared filter, and orthomode transducer (OMT) can show a standing-wave-like structure in the instrumental polarization \citep{Gonzalez2014}.  In addition, the infrared (IR) filter in front of the horn causes a strong resonance in the cross polarizations at certain frequencies.  Since the ALMA Band 6 receivers have IR filters and use an OMT for polarization separation, those effects can apply to these Band 6 data.  

\subsection{A.1.2 Reproducibility of the Instrumental Polarization Calibration}
\begin{figure}
\begin{center}
\includegraphics[width=16cm]{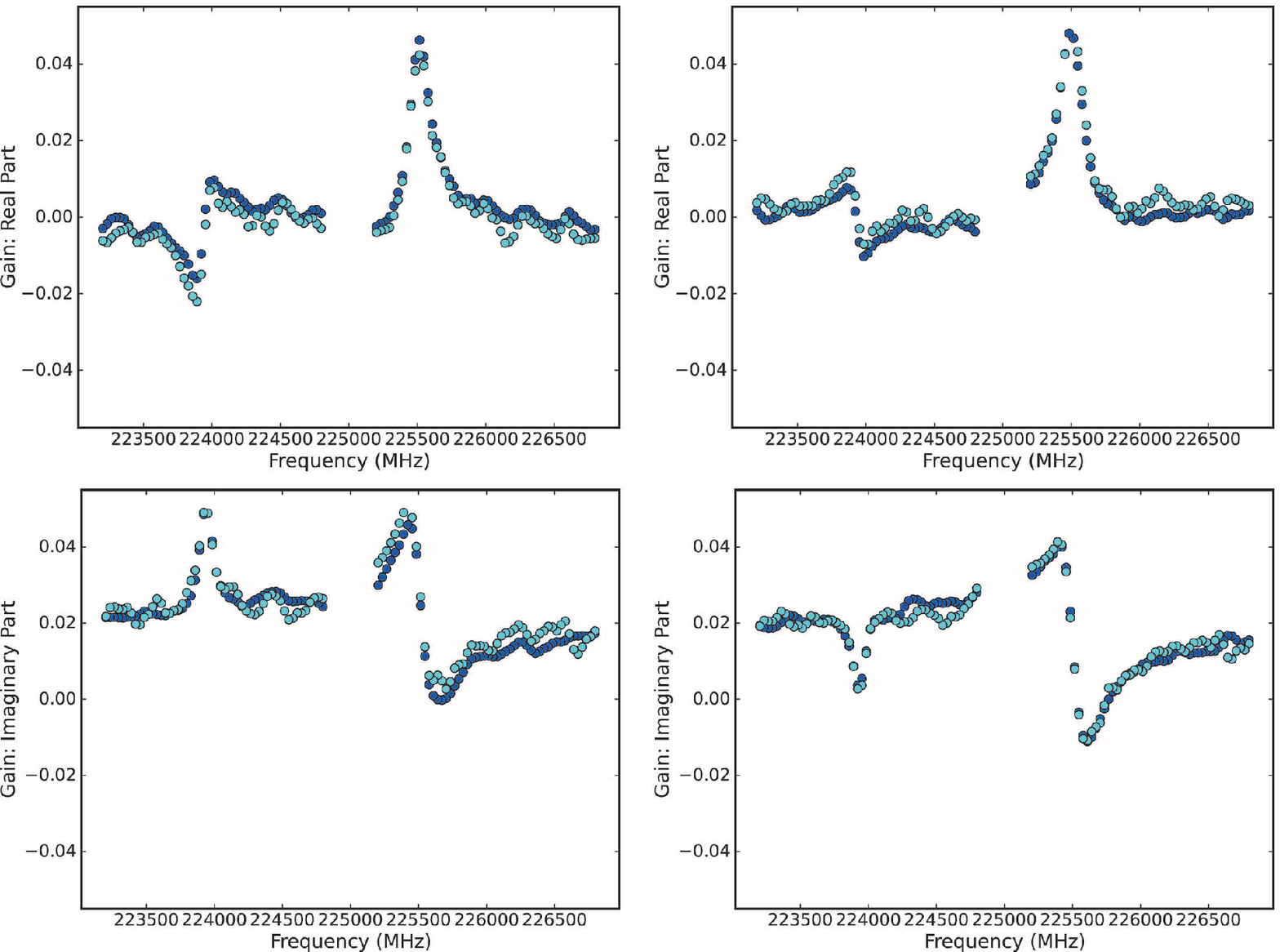}
\end{center}
\caption{The comparison of the $D$-term spectra obtained from J1337-1257 and 3C~279 for antenna DV07. The left and right columns show the $D$-term spectra of $D_\textrm{X}$ and $D_\textrm{Y}$, respectively.  The upper and lower panels show the real and imaginary parts of the $D$-terms.  Blue and cyan symbols represent the $D$-term solutions obtained from 3C~279 and J1337-1257.}
\label{fig:Dcomp}
\end{figure}

For a consistency check, we also performed the polarization calibration using 3C~279, which was originally intended for bandpass calibration.  3C~279 was observed three times at different parallactic angles, allowing us to solve for the instrumental polarization.  The resultant instrumental polarization spectra are consistent with those obtained from J1337-1257.  Although the global consistency of instrumental polarization is confirmed, there are slight discrepancies.  The differences between the two solutions is not greater than 0.5\% and is typically 0.1--0.3\% over the band (Figure \ref{fig:Dcomp}).  Those small discrepancies could result from differing parallactic angles coverage of J1337-1257 and 3C~279, time variation of the instrumental polarization, pointing-direction dependence of the instrumental polarization, and/or other calibration errors.  In any case, these discrepancies can be regarded as the calibration error of the instrumental polarization.  We note that these errors are largely independent of frequency and antenna, and thus they are reduced by averaging over frequencies and antennas.  The measured image rms in the Stokes $Q$ and $U$ images is $\sim$\,$10^{-4}$ of the peak Stokes $I$ intensity.  Thus, the data demonstrates a detectability of linear polarization at the $<0.1$\% level.  
 
\subsection{A.1.3 Calibration Error of Instrumental Polarization}
One of the major sources of error that limits the sensitivity of polarization images is error in the measurement of the instrumental polarization on each antenna.  Here we provide an approximate error estimate.  The cross correlations for linear feeds on a baseline between antenna \textit{ i} and antenna \textit{ j} is given by
\begin{eqnarray}
X_{i}X_{j}^{*} &=& (I+Q_{\psi})+U_{\psi}(D^{*}_{X_{j}}+ D_{X_{i}})  \\
X_{i}Y_{j}^{*} &=& U_{\psi}+I(D^{*}_{Y_{j}}+ D_{X_{i}})+Q_{\psi}( D^{*}_{Y_{j}}-D_{X_{i}}) \\
Y_{i}X_{j}^{*} &=& U_{\psi}+I(D_{Y_{i}}+ D^{*}_{X_{j}})+Q_{\psi}( D_{Y_{i}}-D^{*}_{X_{j}}) \\
Y_{i}Y_{j}^{*} &=& (I-Q_{\psi})+U_{\psi}(D^{*}_{X_{j}}+ D_{X_{i}})\,\,, 
\end{eqnarray}
where $Q_{\psi}=Q\cos 2\psi +U\sin 2\psi$, $U_{\psi}=-Q\cos 2\psi +U\sin 2\psi$, $\psi$ is parallactic angle, and $D_{X}$ and $D_{Y}$ are the instrumental polarization $D$-terms.  Here we assume that the second-order $D$-terms are negligible since the first-order $D$-term level is typically 2--3\%.  We also assume that the Stokes V is negligible for simplicity.  Each Stokes parameter can be obtained from these four equations, and thus the calibration residuals of $D_{X_{i}}$, $D_{X_{j}}$, $D_{Y_{i}}$, $D_{Y_{j}}$ affect the $Q$ and $U$ visibilities.  Assuming the $D$-term errors are all independent,
\begin{equation}
\sigma_{Q_{ij}}^{2}\simeq\sigma_{U_{ij}}^{2}\simeq \frac{I_{ij}^{2}}{4}(\sigma_{D_{X_{i}}}^{2}+\sigma_{D_{Y_{i}}}^{2}+\sigma_{D_{X_{j}}}^{2}+\sigma_{D_{Y_{j}}}^{2})\,\,,
\end{equation}
where $\sigma_{Q_{ij}}$ is the error of Stokes $Q$ visibility, $\sigma_{U_{ij}}$ is the error of Stokes $U$ visibility, $\sigma_{D_{X_{i}}}$ is the error of $D_{X_{i}}$, $\sigma_{D_{Y_{i}}}$ is the error of $D_{Y_{i}}$, $\sigma_{D_{X_{j}}}$ is the error of $D_{X_{j}}$, and $\sigma_{D_{Y_{j}}}$ is the error of $D_{Y_{j}}$.  If the errors in $D$-terms are all about the same ($\sigma_{D}^{2}\equiv\sigma_{D_{X_{i}}}^{2}=\sigma_{D_{X_{j}}}^{2}=\sigma_{D_{Y_{i}}}^{2}=\sigma_{D_{Y_{j}}}^{2}$), then
\begin{equation}
\sigma_{Q_{ij}}^{2}\simeq \sigma_{U_{ij}}^{2}\simeq I_{ij}^{2}\sigma_{D}^{2}\,\,.
\end{equation}
If we assume that all $D$-terms are independent over $N$ antennas, the error in the Stokes $Q$ and $U$ images is
\begin{equation}
\sigma_{Q}\simeq \sigma_{U}\simeq\frac{I\sigma_{D}}{\sqrt{N}}\,\,.
\end{equation}
As we mentioned before, the calibration error of the $D$-terms is typically $\sim$\,$10^{-3}$ for each spw.  About 30 antennas were used during Cycle 2 ALMA observations; thus, we obtain $\sigma_{Q}\simeq \sigma_{U}\simeq0.0002I$.  

As we presented in Section \ref{sec:obs}, the measured rms noise values in the Stokes $Q$ and $U$ images of 3C~286 data are 0.035~mJy beam$^{-1}$ and 0.041~mJy beam$^{-1}$, respectively.  The peak intensity of Stokes $I$ is 388~mJy beam$^{-1}$.  Thus, the measured $\sigma_{Q}$ and $\sigma_{U}$ are $\sim$\,$10^{-4}$, which agrees with above estimation to within a factor of two.

\section{A.2 Core Shift Analysis for 3C~286}
\begin{figure}
\begin{center}
\includegraphics[width=12cm]{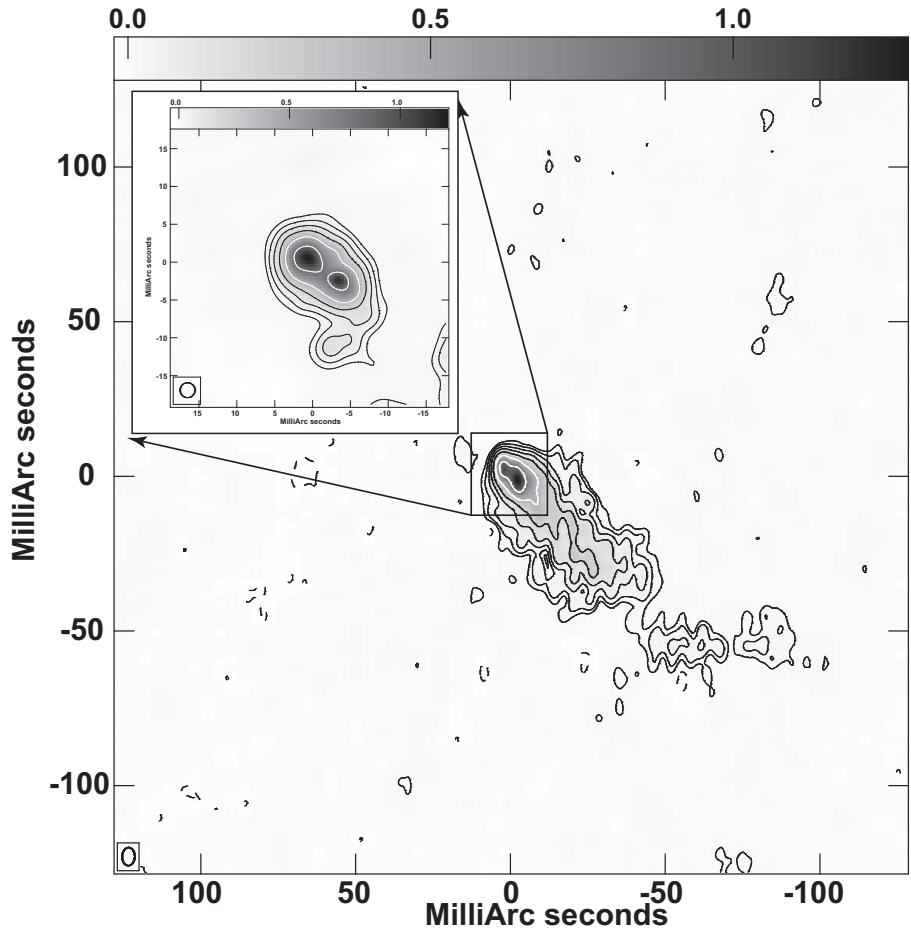}
\end{center}
\caption{VLBI contour image of 3C~286 at 13~cm.  The contours are plotted at 13.1~mJy $\times$ (--1, 1, 2, 4, 8, 16, 32, 64, 128, 256, 512).  The synthesized beam size is $4.34 \arcsec \times 3.59\arcsec$ at a position angle of $-48.5^{\circ}$.  The zoomed image in this figure is the super resolution map of the central region.  The image was produced using a circular Gaussian beam with a FWHM of approximately half of the synthesized beam.}
\label{fig:sbandimage}
\end{figure}

%\subsection{A2.1 Component Identification}
For the core-shift analysis, we performed Gaussian fitting of C1 and C2 in the VLBI image and measured the separation of two components (Table \ref{tab:separation}).  Using this method could introduce position uncertainties due to the possible blending of the core and the jet emission; this would cause the separation to be underestimated \citep[e.g.,][]{Sudo2003, Guirado1995, Hada2011}.  In the 13~cm (2.3~GHz) image in particular, the separation between C1 and C2 is comparable to the beam size ($5.86\arcsec \times 4.05\arcsec$ at a position angle of $-1.95^{\circ}$ for the data taken on 1997 January 11 and $4.34 \times 3.59\arcsec$ at a position angle of $-48.5^{\circ}$ for the data taken on 2000 October 23).  To reduce the blending effect, we produced a super resolution image (Figure \ref{fig:sbandimage}) by using a circular Gaussian beam with a full-width at half-maximum (FWHM) of about half of the synthesized beam.  We then defined the peak position of the components derived from the Gaussian fitting as the core position.  With this technique, C1 and C2 are well separated and we can extract the position of each component without blending.  The separation of the two components was 5.2~mas (1997 January 11) and 4.8~mas (2000 October 23) in the super resolution images, whereas it was 4.7~mas (1997 January 11) and 4.3~mas (2000 October 23) in the original images.  We adopt the separations derived from the super resolution image as the separation of two component at 13~cm.  We also applied the same method to the 4~cm and 2~cm images, but the resultant change in separation between the two methods was $<0.1$~mas.  This is because the two components are already well-separated in those images, which were created with the original synthesized beam size.  We thus use the original beam size to define the separation of the two components in the 4~cm and 2~cm images.  As a further check, we made 4~cm images convolved with circular Gaussians with FWHMs of 1.2, 1.5, and 2.0 $\times$ the geometric mean of synthesized beam.  We confirmed that the resultant separation monotonically decreased from $\sim$\,5.5~mas to $\sim$\,4.5~mas with increasing size of the convolving beam, as expected.

\end{document}